\documentclass[twocolumn,aps,prb,showpacs]{revtex4}
\usepackage{amsmath}
\usepackage{epsfig}
\usepackage{epsf}
\usepackage{array}
\usepackage{color}
\usepackage{graphicx}
\usepackage{epstopdf}

\begin{document}
\bibliographystyle{apsrev}

\title{Optical conductivity in cluster dynamical mean field theory: formalism and application to high temperature superconductors }
\author{Nan Lin, Emanuel Gull and A. J. Millis}
\affiliation{Department of Physics, Columbia University, 538 West 120th Street, New York, New York 10027, USA}
\date{\today }

\begin{abstract}
The optical conductivity of the one-band Hubbard model  is calculated using  the `Dynamical Cluster Approximation' implementation of dynamical mean field theory  for parameters  appropriate to high temperature copper-oxide superconductors. The calculation includes vertex corrections and the result demonstrates their importance.  At densities of one electron per site, an insulating state is found with gap value and   above-gap absorption  consistent with measurements. As carriers are added the above gap conductivity rapidly weakens and a three component structure emerges, with a low frequency `Drude' peak, a mid-infrared absorption, and a remnant of the insulating gap. The mid-infrared feature obtained at intermediate dopings  is shown to arise from a pseudogap structure in the density of states. On further doping the conductivity evolves to the Drude peak plus weakly frequency dependent tail structure expected for less strongly correlated metals.
 \end{abstract}


\maketitle

The frequency-dependent (`optical') conductivity $\sigma(\Omega)$ is an important probe of  electronic condensed matter physics, revealing electronic band gaps, scattering processes, carrier number and effective mass. Optical conductivity measurements have played a particularly important role in the study of high temperature copper-oxide superconductors \cite{Uchida91,Basov05}, revealing behavior which deviates sharply from conventional band theoretic expectations.  In the undoped `parent compounds' such as $La_2CuO_4$ measurements \cite{Uchida91} reveal a large $\sim 1.75eV$ gap which persists essentially unchanged  as temperature is raised above the N\'{e}el temperature $T_N\sim 300K$ while band theory predicts metallic behavior in the absence of antiferromagnetism. As the materials are doped, the gap feature weakens and absorption  appears at lower frequencies. The lower frequency absorption is  often \cite{Cooper90}  decomposed into two parts, a `Drude' peak centered at $\Omega=0$ and a `mid-IR' structure  at $\Omega \sim 0.5eV$; both parts are characterized by an oscillator strength (integral of conductivity over a frequency range) which is small compared to the band theory value and increases as the doping is increased \cite{Comanac08}. The `mid-IR' band has been variously interpreted as an effect of scattering of carriers from spin fluctuations \cite{Zemlji05,Abanov04} or other bosons \cite{Varma89,Timusk91}, a signature  of two-component absorption \cite{Tanner95} and as an indication of a novel charge 2e excitation \cite{Choy08,Chakraborty08}, but a clear consensus on the interpretation has not emerged.

\begin{figure}[t]
\includegraphics[width=0.8\columnwidth]{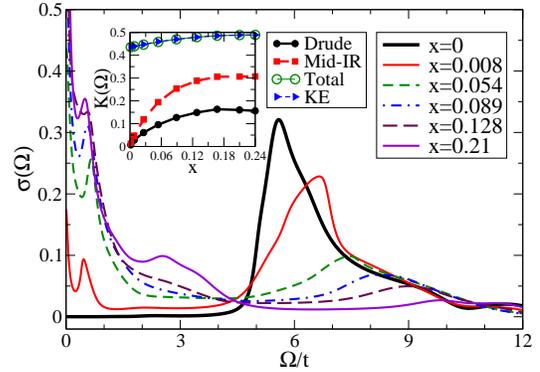}
\caption{Optical conductivity calculated for  indicated dopings using 4-site DCA approximation to the paramagnetic phase of the two-dimensional one-band Hubbard model at parameters corresponding to high temperature superconductors at temperature $T\approx 400K$ and other parameters described in the text. Inset: $2/\pi$ times integral of optical conductivity over low frequency $\Omega<0.6t$ (`Drude', solid line, black on-line), intermediate frequency $\Omega=2t$ (`Mid-IR', long-dashed  line, red on-line), and all frequencies (`Total', light solid line, green on-line), along with independently computed kinetic energy (`KE', light dashed line, blue on-line).} 
\label{dopingfig}
\end{figure}

Cluster dynamical mean field approximations \cite{Hettler98,Kotliar01} to the one-band Hubbard model have been argued to provide a reasonable description of the physics of the cuprates. Single electron properties such as the photoemission spectra have been argued \cite{Civelli05,Macridin06,8SitePaper,Gull08plaquette,Park08plaquette} to be in good agreement with data. However, in cluster dynamical mean field theory evaluation of two particle response functions such  the optical conductivity requires computation  of a vertex function. In this paper we show that the vertex correction may be computed and makes a significant contribution (especially to the conductivity of  undoped and lightly doped materials). 
Our main  result, a computation of the  variation with doping of the frequency dependent conductivity $\sigma(\Omega)$  of the Hubbard model at physically relevant parameters, is shown in Fig.~\ref{dopingfig}. The curves bear a striking similarity to the conductivity of hole-doped high temperature superconductors \cite{Uchida91,Basov05}. The conductivity calculated for the undoped system displays a characteristic insulating spectrum with a gap $\sim 5t\approx 1.8eV$ similar in magnitude to that observed in $La_2CuO_4$ and an above-gap absorption strength corresponding to $\sim 1300 \Omega^{-1}cm^{-1}$ about $30\%$ larger than observed. On doping, this gap is rapidly destabilized, again in a manner consistent with measurements. Absorption in the near-gap region is suppressed, while substantial absorption strength appears at low frequencies.  As the doping is increased the optical response may be described in terms of  a `Drude' peak indicating coherent quasiparticle motion, an additional  `mid-IR' feature ($\omega \sim 0.5t$)  and a high frequency tail.  By the highest doping the conductivity has evolved to the `Drude peak' plus weak high frequency tail characteristic of Fermi-liquid metals.  

To further characterize the evolution of the conductivity we present in the inset of Fig.~\ref{dopingfig} the partial  optical integrals $K(\Omega)=\frac{2}{\pi}\int_0^\Omega d\omega \sigma(\omega)$  for a low frequency ($\Omega=0.6t$ encompassing the Drude peak), an intermediate frequency $\Omega=2t$ , somewhat less than  half of the gap value  and the total integral $\Omega\rightarrow \infty$ (which is seen to agree with the independently calculated kinetic energy).   The shift of spectral weight, first into a mid-IR band and then into a Drude peak is very similar to observations in high-$T_c$ cuprates,  although the calculated mid-IR spectral weight is higher than values inferred from data (see e.g. Fig.~1 in Refs.~\onlinecite{Uchida91,Comanac08}.)  

The derivation of our results begins from the current-current response function $\chi_{jj}(t)=i\partial_t \sigma(t)$ which relates  a spatially uniform, time dependent (therefore transverse) electric field ${\vec E}$ to the current ${\vec j}$ it creates. For a system described by a Hamiltonian ${\mathbf H}={\mathbf T}+{\mathbf U}$ with  interactions $\mathbf{U}$ which depend only on particle and spin densities (not on particle or spin  currents) we have ${\vec j}(t)=\text{Tr}[\vec{\mathbf {J}}(A){\mathbf G}(t;\{{\vec {A}}\})]$. The current operator $\vec{\mathbf{J}}$ is obtained from the derivative of the single-particle terms $\mathbf{T}$ with respect to vector potential:  ${\vec {\mathbf {J}}}=\delta {\mathbf T}/\delta {\vec A}$ while the electron Green function ${\mathbf G}(t;\{\vec{A}\})=(i\partial_t-{\mathbf T}(\{{\vec A}\})-{\mathbf \Sigma}(\{{\vec A}\}))^{-1}$ is to be computed in the presence of the time dependent vector potential ${\vec A}$.  Here bold face quantities denote matrices in the space of states of the system (including the spatial indices). Expanding to linear order in ${\vec A}$ and introducing  the  `kinetic energy' operator $\mathbf {K}=\delta^2 {\mathbf{ T}}/\delta {\vec A}^2$ and vertex operator ${\vec{\mathbf {\Gamma}}}=\delta{\mathbf{\Sigma}}/\delta{\vec A}$  we obtain $\chi_{jj}=\chi_\text{dia}+\chi_\text{bubble}+\chi_\text{vertex} $ with
\begin{align}
\chi_\text{dia}(t-t')&=\text{Tr}\left[{\mathbf {K}}{\mathbf G}(t=0)\right]\delta(t-t'),
\label{chidia} \\
\chi_\text{bubble}(t-t')&=\text{Tr}\left[\vec{{\mathbf {J}}}{\mathbf G}(t-t')\vec{{\mathbf {J}}}{\mathbf G}(t'-t)\right],
\label{chibubble} \\
\chi_\text{vertex}(t-t')&=\text{Tr}\left[\vec{{\mathbf {J}}}{\mathbf G}(t-t_1)\vec{{\mathbf {\Gamma}}}(t_1-t',t'-t_2){\mathbf G}(t_2-t) \right],
\label{chivertex}
\end{align} 
(convolution on internal time indices is to be understood).
Combining the Kramers-Kronig relation between real $(^{'})$ and imaginary $(^{''})$ parts of $\sigma$ with the gauge-invariance condition that for a non-superconducting material  $\chi_{jj}(\Omega=0)=0$   implies \cite{Kohn64} $\chi_\text{dia}=\int\frac{d\omega}{\pi} \chi_{jj}^{''}(\omega)/\omega=\int \frac{d\omega}{\pi}\sigma^{'}(\omega)$, which is the usual f-sum rule.

In the dynamical cluster approximation (DCA)\cite{Hettler98,Maier05}  implementation of the dynamical mean field approximation \cite{Georges96}  one tiles the Brillouin zone into $a=1...N$ non-overlapping equal-area regions and approximates the electron self-energy $\Sigma(k,\omega)$ by the piecewise constant form
\begin{equation}
\Sigma(k,\omega)=\sum_a^N\Sigma_a(\omega)\phi_a(k)
\label{sigmadca}
\end{equation}
with $\phi_a(k)=1$ if $k$ is in tile $a$ and zero otherwise.  Thus if $N\neq 1$ the self-energy has an explicit momentum dependence arising from the discontinuities at the boundaries of the tiles.  The self energies $\Sigma_a$ are  computed from the solution of an $N$-site  quantum impurity model which involves the interactions of the original model (projected onto the impurity cluster) and  mean field functions ${\cal G}_a^{-1}$ which are  fixed by the self consistency equations
\begin{equation}
{\cal G}_a^{-1}=\Sigma_a+\left[\int _a(dk)G(k)\right]^{-1}
\label{dcasce}
\end{equation}
where  the integral is over the tile $a$ with appropriate measure $(dk)$, $G(k)$ is the Green function of the lattice problem computed with $\Sigma$ defined by Eq.~(\ref{sigmadca}).

For the conductivity we require the vertex function $\vec \Gamma(\omega+\Omega,\omega)\equiv \delta \Sigma/\delta {\vec A}$, which has two sources: the explicit dependence on $k$ arising from the momentum-space discontinuities and any additional dependence of $\Sigma$ on ${\vec A}$.  The additional ${\vec A}$-dependence  arises via the impurity model from  a dependence of ${\cal G}^{-1}$ on $A$ which may be computed by linearizing Eq.~(\ref{dcasce}) in $A$.  Denoting the first order changes in ${\cal G}$ and $\Sigma$ by $\delta {\cal G}$ and $\delta \Sigma$ we have
\begin{equation}
\delta{\cal G}^{-1}_\alpha -I_\Sigma[\{\delta \Sigma_\alpha\}]=-{\vec I}^\alpha_v\cdot {\vec A}
\label{lineardcasce}
\end{equation}
with (time arguments are not written explicitly)
\begin{eqnarray}
{\vec I}_v&=&- G^{-1}_a\left(\int_a(dk)G(k)\frac{\partial \varepsilon}{\partial {\vec k}} G(k) \right) G^{-1}_a
\label{Iv}\\
I_\Sigma&=&\delta \Sigma^a +G^{-1}_a\left(\int_\alpha(dk)G(k)\delta
		\Sigma^\alpha  G(k) \right)G^{-1}_a
\label{Isigma}
\end{eqnarray}

The canonical tiling for clusters of size   $N=1,2,4$  produces momentum sectors with symmetry  such that $I_v=0$. Thus as noted by Ref.~\onlinecite{Jarrell95} (for $N=1$) and \onlinecite{Haule06} (for $N=4$), in these clusters there is no explicit dependence of $\Sigma_a$ on $A$. Ref.~\onlinecite{Haule06} further argued that  for $N=4$ all vertex corrections vanished. This is incorrect, although the vertex corrections turn out to be unimportant for the situation of interest to Ref.~\onlinecite{Haule06}. The explicit momentum dependence provides a non-vanishing vertex correction arising from the momentum space discontinuities which  occur along the lines ${\vec k}^{ab}$ separating tile b (on the of larger k side) from tile a (on the smaller k side).  We define ${\vec n}^{ab}$ to be the normal to this line. In the  $\Omega=0$ limit the  vertex correction is directly given by $\partial \Sigma/\partial {\vec k}$. To determine the vertex correction for $\Omega\neq 0$ we consider the Ward identity $\Omega \Gamma_\rho-\vec{q}\cdot{\vec \Gamma}_J=G^{-1}(k+q,\omega+\Omega)-G^{-1}(k,\omega)$ relating the charge $\Gamma_\rho$ and current ${\vec \Gamma}_J$ vertices to the inverse Green functions. Within the DCA approximation $\Gamma_\rho$ arises from the functional derivative of $\Sigma$ with respect to a time dependent chemical potential; it is computed along the lines of Eq.~(\ref{lineardcasce}) but with $\partial \varepsilon_k/\partial {\vec k} \cdot {\vec A}$ replaced by $\delta \mu$. Because this perturbation is a scalar it has no contribution from the functions $\phi_k$ and therefore cannot have any term proportional to a delta function of $k$. Thus we may identify the current vertex from the contributions in the Ward identity proportional to delta functions in $k$-space, yielding
\begin{equation}
{\vec \Gamma}^{k}(\omega+\Omega,\omega)={\vec n}^{ab}\left(\Sigma_b(\omega+\Omega)-\Sigma_a(\omega)\right)\delta\left(({\vec k}-{\vec k}^{ab})\cdot {\vec n}^{ab}\right)
\label{gammadynamic}
\end{equation}

We apply the formalism to the $4$ site DCA approximation to the two dimensional Hubbard model, $H=\sum_{k,\sigma}\varepsilon_kc^\dagger_{k,\sigma}c_{k\sigma}+U\sum_in_{i\uparrow}n_{i\downarrow}$ with  $\varepsilon_k=-2t\left(\cos k_x+\cos k_y\right)-4t'\cos k_x\cos k_y$. The current operator for the $x$ direction is $j_x=2t\sin k_x+4t'\sin k_x\cos k_y$ and the kinetic energy operator $K=2t\cos k_x+4t'\cos k_x\cos k_y$.  For this approximation, Eq.~\ref{gammadynamic} is the appropriate vertex correction. We restrict attention to the paramagnetic phase. We use the numerically exact continuous-time auxiliary field\cite{ct-aux} (CT-AUX) impurity solver to solve the impurity model and construct vertex functions and the conductivity. Parameters relevant to high temperature copper-oxide superconductors are $t\approx 0.35eV$, $t'\approx -0.3 t$ \cite{Andersen95} and $U\sim 9t$ \cite{Comanac08}.  The precision of calculations for $t'\neq 0$ or $n\neq 1$ is limited by a fermion sign problem; we therefore  present results for $U=6t$ and $t'=0$ where higher precision data can be obtained. 

\begin{figure} [t]
\includegraphics[angle=-90, width=0.85\columnwidth]{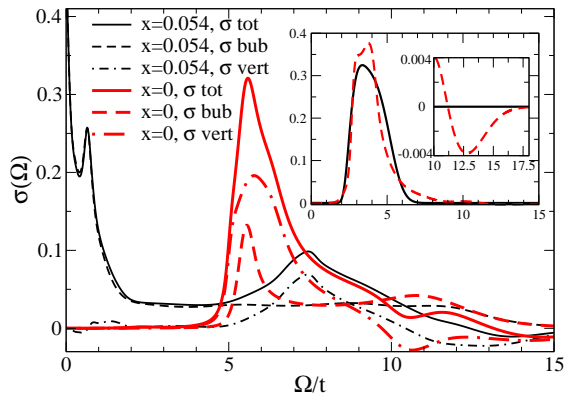}
\caption{Main panel: conductivity calculated from 4-site DCA approximation to one-band Hubbard model  for $U=9t$, $t'=-0.3t$ and dopings $x=0$ (heavy lines, red on-line) and $0.054$ (light lines, black on-line) by continuing self energy. Full lines: conductivity; dashed lines: contribution to conductivity from convolution of bubble diagrams Eq.~\ref{chibubble}; dash dotted lines: contribution from vertex corrections, Eq.~\ref{chivertex}. Larger Inset: comparison of conductivity computed by continuing self energy  (dashed line, red on-line) and continuing Matsubara response function (solid line, black on-line) at $U=6t$ and $t'=0$. Smaller inset: expansion of high frequency negative conductivity region for $U=6t$ and $t'=0$.  }
\label{halffilledfig}
\end{figure}

Because we solve the model on the imaginary axis, an analytical continuation is required to obtained real frequency information. This may be done in two ways: either by continuing the self energies\cite{Wang09a} and then computing the conductivity or by continuing directly the Matsubara-axis conductivity. Analytical continuation requires a covariance matrix of error estimates. For the self energy the covariance matrix is available from the QMC data, while for the conductivity we estimate the covariance (which is highly non-diagonal because of the $1/\Omega$ in the definition) from an ensemble of $16$ independent QMC solutions of the mean field equations. The continuations are tested by comparing the directly computed Matsubara axis $\sigma(i\Omega_n)$ to the same quantity `back-computed'  from the continuation. Significant differences in real-axis conductivity correspond to variations of a few times $10^{-4}$ in $\sigma(i\Omega_n)$, setting a stringent requirement on the quality of the data.  The inset  of Fig.~\ref{halffilledfig} presents highly precise data obtained at the sign-problem-free parameters $U=6t$ and $t'=0$; we see that the  two continuation methods yield similar results; differences between them are a measure of the best-case uncertainties in the continued $\sigma$. At $U=9t$ and $t'=-0.3t$ the differences between the methods are larger; in particular the gap edges are much broader in the traces obtained by continuing the Matsubara axis response functions; we believe this broadening is unphysical for the reasons given in Ref.~\onlinecite{Wang09a}.  The  conductivity obtained from continuing $\sigma(i\Omega_n)$ is found in general to produce a back-continued $\sigma$ in worse agreement with original data than the conductivity obtained by continuing the self energy; we therefore present results obtained from the latter method. One difficulty must be noted. As can be seen from Fig.~\ref{halffilledfig}, in the high frequency regime (frequencies well above the gap) the vertex correction acts to cut off the high frequency tail found in $\chi_{\text{bubble}}$, adding  a negative contribution to the positive-definite $\chi_{\text{bubble}}$ so that the total contribution nearly vanishes. In calculations based on continuing the self energy first, the vertex correction in fact overcompensates,  leading to an unphysical negative conductivity for some high frequencies  (if $\sigma(i\Omega_n)$ is directly continued the $\sigma^{'}(\omega)$ is by construction positive).  We  believe that the overcompensation is a numerical artifact.  Our numerical uncertainties, both from the QMC measurement and the analytical continuation, are largest in this regime. The magnitude of the unphysical negative contribution is small: for our $U=9t$ calculations the spectral weight in the negative region ranges from $5\%$ of the total spectral weight at $x=0$ to $0.4\%$ at our highest doping.   Further, the overshoot  is found to decrease as the numerical accuracy of our computation is improved and as seen from the second inset to Fig.~\ref{halffilledfig}   can  be made  smaller than $2\%$ of the total spectral weight in our best case. However,  to date we have been unable to eliminate the negative region entirely.  

The main panel  of Fig.~\ref{halffilledfig} presents the conductivity for $t'=-0.3t$ and $U=9t$ as well as its decomposition into `bubble' (Eq.~\ref{chibubble}) and `vertex' (Eq.~\ref{chivertex}) contributions for dopings $x=0$ and $x=0.054$.   The vertex correction is seen to make a non-negligible contribution to the conductivity and to be essential to fulfilling the $f$-sum rule.  It decreases in importance as doping increases, as expected because with increased doping the self energy becomes more isotropic in momentum space. This aspect of our result disagrees with Ref.~\onlinecite{Chakraborty08} which stated (on the basis of a comparison of a 4-site CDMFT conductivity computed without vertex corrections to the f-sum rule) that vertex corrections were unimportant near half filling and increased in importance as the doping increased; on the other hand at lower frequencies and higher dopings where vertex corrections are of less importance our result is similar to that of Ref.~\onlinecite{Chakraborty08}. The origin of the difference is not clear. 

At $x=0$ the vertex corrections bring two effects: they increase the magnitude of the conductivity in the above-gap region and they steepen the rate at which the conductivity rises above the gap edge. We believe that these two effects are consequences of the short ranged order captured by the DCA approximation.  Inspection of the Green function (not shown here) indicates that on the single-particle level the state is an indirect-gap insulator with the highest energy filled states at a different momentum from the lowest energy empty states.  As shown e.g.  in Ref.~\onlinecite{Comanac08} and \onlinecite{Wang09a}, the backfolding associated with long-ranged order converts the indirect gap to a direct one, dramatically steepening the conductivity onset.  The vertex correction provides a similar effect. The  vertex correction also expresses the  `coherence factor' physics associated with strong two-sublattice spatial correlations. To see this, consider a mean-field model of a material with two-sublattice order, for which the electron propagator has `normal' ($G \sim \langle c_kc^\dagger_k\rangle$) and `anomalous' ($F\sim \langle c_kc^\dagger_{k+Q}\rangle$) parts given by  $G(k,\omega)=(\omega-\varepsilon_{k+Q})/((\omega-\varepsilon_k)(\omega-\varepsilon_{k+Q})-\Delta^2)=1/(\omega-\varepsilon_k-\Delta^2/(\omega-\varepsilon_{k+Q}))$ and $F(k,\omega)=\Delta^2/((\omega-\varepsilon_k)(\omega-\varepsilon_{k+Q})-\Delta^2)$ respectively.  The conductivity in the ordered state is computed from the sum of a `$G-G$' and an `$F-F$' bubble which give equal contributions to the conductivity for frequencies near the gap edge. The state uncovered in the 4-site DCA calculation has no long ranged order, so the anomalous ($F$) part vanishes and convolution of bubble diagrams would  capture only the $G-G$ contribution. The vertex corrections in effect add back the $F-F$ term. 

\begin{figure}[t]
\includegraphics[angle=-90,width=0.85\columnwidth]{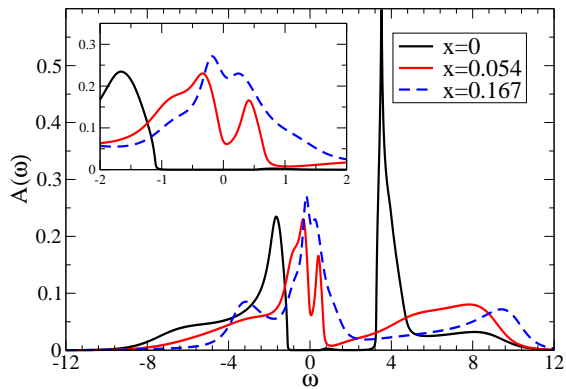}
\caption{Main panel: Density of states calculated for the sector containing the Fermi surface from 4-site DCA approximation at $U=9t$ and $t'=-0.3t$ at dopings indicated. Inset: expansion of the near-Fermi-surface region.}
\label{dosfig}
\end{figure}

The conductivity is related to the electron spectral function, shown in Fig.~\ref{dosfig}.  The initial doping moves the chemical potential into the lower Hubbard band and rapidly broadens the sharp peak at the edge of the upper Hubbard band; we see also from Fig.~\ref{dopingfig} that the form of the near-gap-edge conductivity changes substantially.  Interestingly, in the doped materials  the remains of the above-gap  absorption is entirely expressed by the vertex correction.  At intermediate dopings $x=0.054$ and $0.089$ the many-body density of states exhibits a `pseudogap', a small gap at the Fermi level previously noted  \cite{Civelli05,Gull08plaquette,Ferrero09}. Excitations across the pseudogap have the correct energy to account for the mid-IR feature observed in the data and in high-$T_c$ materials (a similar connection was made in Ref \onlinecite{Choy08,Chakraborty08}).  

To summarize, we have presented theoretically consistent calculations of the optical conductivity of the Hubbard model within the `DCA' implementation of cluster dynamical mean field theory.  The calculated results bear a very great similarity to the conductivity observed in high temperature superconductors. The important role played by spatial correlations in the cluster DMFT approximation (expressed in the calculation by vertex corrections) is seen from the rapid rise of the conductivity above the gap edge and the rapid changes with doping, while vertex corrections are less important at higher doping and lower frequency. Important directions for future study include extensions to the case of Raman scattering and to larger clusters.

{\it Acknowledgments:} We thank P. Philips, Jie Lin and X. Wang for helpful conversations and  acknowledge support from the National Science Foundation Division of Materials Research under grant DMR-0705847. QMC calculations have been performed on the Brutus cluster at ETH Zurich, using a code based on ALPS\cite{ALPS}.


\end{document}